# Complex Networks *in* and *beyond* Physics

**Physics** (Greek: *φύσις* [phúsis], nature) is the branch of science concerned with the characterization of universal laws of Nature portraying its logically ordered picture in agreement with experience. Theoretical physics is closely related to mathematics, which provides a language for physical theories and allows for a rationalization of thought formulating these laws in terms of mathematical relations.

Physicists study a wide variety of phenomena creating new *interdisciplinary* research fields by applying theories and methods originally developed in physics in order to solve problems in economics, social science, biology, medicine, technology, etc. In their turn, these different branches of science inspire the invention of new concepts in physics. A basic tool of analysis, in such a context, is the mathematical theory of *complexity* concerned with the study of *complex systems* including human economies, climate, nervous systems, cells and living things, including human beings, as well as modern energy or communication infrastructures which are all *networks* of some kind. Recently, complexity has become a natural domain of interest of the real world *socio-cognitive* systems, *linguistics,* and emerging *systemics*[1] research. The phenomena to be studied and understood arise from neither the physical laws nor the abstraction of mathematics. The challenge is to discern and formulate plausible *mathematical structures* to describe problems that represent vague human goals. In lack of permanent collaboration with linguists, it is unthinkable to translate problems of considerable difficulty into the language of theoretical physics and graph theory, after to find appropriate models and finally to interpret the results.

The interest of this project is manifold but particularly by its interdisciplinary character, where scientists coming from *linguistics* and *theoretical physics* interact and contribute with their different approaches and view points to the understanding of very complex issues.

Complex systems appear as a result of the interplay between *Topology* determined by a connected graph, *Dynamics* described by the operators invariant with respect to graph symmetry, and properties of *Embedding* (Euclidean) space specified by a set of measures and weights assigned to elements of the graph [1]. A great deal of research in theory of complex systems has been performed in our team in the last two decades.

In the context of *complex networks* theory created by physicists, the non-trivial topological structure of large networks is investigated by means of various *statistical distributions*. The structure and the properties of complex networks essentially depend on the way how nodes get connected to each other. In most of complex networks emerging in society and technology, each node has a feature which *attracts* the others. In a class of simple models [2], the network dynamics can be described in terms of property of the node and the affinity other nodes have towards that property (*Cameo graphs*). Networks built according to this principle have a degree distribution with a power law tail, whose exponent is determined only by the nodes with the largest affinity values. It appears that the *extremists* lead the formation process of the network and manage to shape the final topology of the system.

---

[1] **Systemics** is defined as an emerging branch of science that studies holistic systems and tries to develop logical, mathematical, engineering and philosophical paradigms and frameworks in which physical, technological, biological, social, cognitive and metaphysical systems can be studied and developed.

The *exceptional events* play a crucial role in the formation of network structures [9]. The dynamics of some vertices — the "*hubs*" which have an extremely high number of connections to other vertices is of primary importance for complex networks. These networks are generally "*scale-free*"; in other words, they exhibit architectural and statistical *stability* as the degree distribution grows. A class of probabilistic model for a system at a threshold of *instability* has been studied in [4]. The distribution of residence times below the threshold characterizes the properties of such a system. Being at a threshold of instability, the system can induce various types of random graphs and the *scale free random graphs* among others [5]. The priority-based scheduling rules in single-stage queuing systems (QS) generate *fat tail* behavior for the task waiting time distributions induced by the waiting times of very low priority tasks that stay unserved almost forever as the task priority indices are "frozen in time" [6]. The task waiting time distributions have been studied for a population-type model with an age structure and a QS with deadlines assigned to the incoming tasks, which is operated under the "earliest-deadline-first" policy. As the aging mechanism ultimately assigns high priority to any long waiting tasks, fat tails cannot find their origin in the scheduling rule alone.

Graphs obtained by successive creation and elimination of edges into small neighborhoods of the vertices evolve towards *small world graphs* with logarithmic diameter, high *clustering coefficients* and a fat tail distribution for the degree [7]. It is important to note that it was only local edge formation processes that rise small worlds, no preferential attachment was used. Simple edge generation rules based on an inverse like mass action principle for random graphs over a structured vertex set, under very weak assumptions on the structure generating distribution, also yield a scale free distribution for the degree [8]. A local search principle important in many social applications, "*my friends are your friends*" have also been introduced and studied; networks generated in accordance to such a principle have essentially high clustering coefficients.

Although investigations into the statistical properties of graphs such as a heavy-tail in the degree distribution of nodes could uncover their hierarchical structure, they are futile if the detailed information on the structure of graphs is of primary interest since many graphs characterized by similar statistics of node degrees and shortest path lengths can be of dramatically different structures [1]. The structure and symmetry of graphs play the crucial role in behavior of dynamical systems defined on that. It can be clearly demonstrated in epidemiological research describing the dynamics of sexually transmitted diseases, the Human Immune Deficiency Virus (HIV) and AIDS, in particular [9]. Mathematical modeling on the spread of sexually transmitted diseases [10,11,12,13,14,15] studied on various random graphs displays the importance of critical parameters such as the transmission probability and edge creation probability for the epidemic spreading. It has been found that the epidemic spreading in scale-free networks is very sensitive to the *statistics of degree* distribution, the *effective spreading rate*, the *social strategy* used by individuals to choose a partner, and the *policy of administrating* a cure to an infected node [16]. Depending on the interplay of these four factors, the stationary fractions of infected population as well as the epidemic threshold properties can be essentially different. For a model of scale-free graphs with *biased partner choice* that knowing the exponent for the degree distribution is in general *not sufficient* to decide epidemic threshold properties for exponents less than three [17]. Absence of epidemic threshold happens precisely when a positive fraction of the nodes form a cluster of bounded diameter. Probably, it is impossible to obtain a simple immunization program that can be simultaneously effective for all types of scale-free networks [16].

A similar approach can be applied in order to study social diseases like c*orruption*. It has been investigated in [18] as a generalized epidemic process on the graph of social relationships. Corruption is characterized by a strong nonlinear dependence of the transmission probability from the local density of corruption and the mean field influence of the overall corruption in the society. Network clustering and the degree-degree correlation play an essential role in these types of dynamics. In particular, it follows that strongly hierarchically organized societies *are more vulnerable* to corruption than democracies. A similar type of modeling can be applied to other social contagion spreading processes like opinion formation, doping usage, social disorders or innovation dynamics. An agent-based model of factual *communication in social systems*, drawing on concepts from *Luhmann's theory of social systems* has been studied in [19]. The agent communications are defined by the exchange of distinct messages. Message selection is based on the history of the communication and developed within the confines of the problem of double contingency. We have examined the notion of learning in the light of the message-exchange description.

In 1932 Zipf observed that the frequency of English words follows a power law distribution. That is, the word frequency that has rank *n* among all word frequencies is proportional to $1/n^\alpha$, where $\alpha$ is close to 1. Estoup observed the same phenomenon for French language in 1916. In fact Zipf-Estoup'd law holds for other human languages as well as for some artificial ones (e.g. programming languages). Methods of complexity theory have been applied to *linguistics* in [20] in purpose of filtering the *core lexicon* of a language from the *content words*, which tend to be extremely frequent in some texts written in specific genres or by certain authors. Distributions of documents over usage frequencies for such words display long tails representing a bunch of documents in which such words are used in abundance. Collections of such the word "contaminated" documents exhibit a *percolation like phase transition*.

In order to pursue the understanding of dynamical processes defined on various types of graphs, we considered a number of different models. We have started from the investigations in transitions to spatio-temporal intermittency in random network of coupled Chaté–Manneville maps [21]. We have shown that spatiotemporal intermittency occurs for some intervals or windows of the values of the network connectivity, coupling strength, and the local parameter of the map. Within the intermittency windows, the system exhibits periodic and other nontrivial collective behaviors. The detailed behavior depends crucially upon the topology of the random graph spanning the network. We have presented a detailed analysis of the results based on the thermodynamic formalism and random graph theory. The results have been applied in order to explain the emergence, spread, and subside of *panic* in communities constituting of a large number of interacting identical agents exhibiting a simple behavior: being affected with panic, an isolated agent inevitably regains his/her composure in a few time steps. It is the *mutual interaction* between agents that induces the emergency of sustained panic in the community [22].

*Genetic regulatory networks* constitute an important example of dynamical systems defined on graphs. Local dynamics of network nodes exhibits multiple stationary states and oscillations depending crucially upon the global topology of a"maximal" graph (comprising of all possible interactions between genes in the network) [23]. The long time behavior observed in the network defined on the homogeneous"maximal" graphs is featured by the fraction of positive interactions (*activations*) allowed between genes. In networks defined on the inhomogeneous directed graphs depleted in cycles, no oscillations arise in the system even if the negative interactions (*inhibitions*) in between genes present therein in abundance. Local dynamics observed in the inhomogeneous scalable regulatory networks is less sensitive to the choice of initial conditions.

In mathematics, the automorphism groups of a graph are studied. They characterize its symmetries, and are therefore very useful in determining certain of its properties. In particular, the Euclidean metric related to dynamics can be defined on some graphs by means of linear operators remaining invariant under the permutations of nodes and satisfying some conservation properties. These operators describe certain dynamical processes defined on graphs such as *random walks* and *diffusions* [1]. We have studied transport through generalized trees in [24]. Trees contain the simple nodes and super-nodes, either well-structured regular subgraphs or those with many triangles. We observe super-diffusion for the highly connected nodes while it is Brownian for the rest of the nodes. Transport within a super-node is affected by the finite size effects vanishing as $N \to \infty$. For a space of even dimensions, $d = 2, 4, 6...$, the finite size effects break down the perturbation theory at small scales and can be regularized by using a heat-kernel expansion.

Diffusion processes and Laplace operators related to them can be used in order to investigate the structure of networks in the spirit of spectral graph theory. In [25], different models of random walks on the *dual graphs* of compact urban structures are considered. Dual graphs have been widely used in the framework of *space syntax theories* for the analysis of spatial configurations. The general idea is that spaces can be broken down into components, analyzed as networks of choices, and then represented as maps and graphs that describe the relative connectivity and integration of those spaces. From these components it is thought to be possible to quantify and describe how easily *navigable* any space is, useful for the design of museums, airports, hospitals, and other settings where *wayfinding* is a significant issue. Space syntax has also been applied to predict the correlation between spatial layouts and social effects such as *crime, traffic flow, sales per unit area*, etc.

Analysis of access times between streets performed in [25] helps to detect the city modularity. The statistical mechanics approach to the ensembles of lazy random walkers is developed. The complexity of city modularity can be measured by information like parameter which plays the role of an individual fingerprint of Genius loci. Global structural properties of a city can be characterized by the thermodynamic parameters calculated in the random walk problem.

Returning to our opening theme work on networks of the real world is a challenging topic for theoretical physics and applied mathematics characterized by
- Identifying and formulating fundamental structures and properties;
- Defining meaningful mathematical phenomena that represent critical features of the real world;
- Creating effective algorithms that solve the resulting mathematical problems.

The extremely complicated and fuzzy structure of language studying by linguistics reflects the perplex structure of human culture in its evolution. Being a universal communication media, a language conforms to a grammar, or a system of rules, used to manipulate symbols and concepts. In the context of complex systems studies, *a language is an evolving network of interacting words*. Though, it is not easy to define the notion of word interaction in a unique way. During its evolution language self-organizes into a complex structure, in which a *core lexicon* comprising of the *mostly connected words* constitutes the language core (amounting to approximately $5 \cdot 10^3$ words) that does not grow while the total number of words used in the language increases. The precise analysis of their peculiar interactions calls for new methods, in particular, and those of complex systems theory appear to be the most suitable. It is worth to mention that *graph theory* is a primary tool for detecting numerous hidden structures in any graph representing relations in massive data sets. Cameo principle and concepts of Luhmann's theory of social systems have to be extended to take care of linguistic methods as well. Spectral graph theory as applied to the analysis of complex networks seems to be a prominent candidate for a unified mathematical theory describing the sustainable evolution of human languages.